\documentclass[%
 reprint,
nofootinbib,
 amsmath,amssymb,
 aps,
]{revtex4-2}

\usepackage{graphicx}
\usepackage{dcolumn}
\usepackage{bm}

\usepackage{amssymb}
\usepackage{booktabs}
\usepackage{colortbl}
\usepackage[dvipsnames]{xcolor}
\usepackage{pifont}
\usepackage{multirow}
\usepackage{hyperref}

\hypersetup{colorlinks=true,linkcolor=blue,citecolor=blue,urlcolor=blue}

\begin{document}

\preprint{APS/123-QED}

\title{Potential Blind Directions at TeraZ}

\author{Mikael Chala}
\email{mikael.chala@ugr.es}
\author{Juan Carlos Criado}%
 \email{jccriadoalamo@ugr.es}
 \affiliation{Departamento de Física Teórica y del Cosmos, Universidad de Granada, Campus de Fuentenueva, E-18071 Granada, Spain} 

\author{Michael Spannowsky}%
 \email{michael.spannowsky@durham.ac.uk}
\affiliation{%
 Institute for Particle Physics Phenomenology, Department of Physics, Durham University, \\ Durham DH1 3LE, U.K.
}%


\begin{abstract}
The next generation of high-luminosity electron-positron colliders, such as FCC-ee and CEPC operating at the $Z$ pole (TeraZ), is expected to deliver unprecedented precision in electroweak measurements. These precision observables are typically interpreted within the Standard Model Effective Field Theory (SMEFT), offering a powerful tool to constrain new physics. However, the large number of independent SMEFT operators allows for the possibility of blind directions, parameter combinations to which electroweak precision data are largely insensitive. In this work, we demonstrate that such blind directions are not merely an artefact of agnostic effective field theory scans, but arise generically in realistic ultraviolet completions involving multiple heavy fields. We identify several concrete multi-field extensions of the Standard Model whose low-energy SMEFT projections align with known blind subspaces, and show that these persist even after accounting for renormalisation group evolution and finite one-loop matching effects. 
Our analysis shows that TeraZ will set a new benchmark in precision for indirect searches, but fully probing the space of possible ultraviolet physics requires going beyond this stage. Later FCC-ee runs at higher centre-of-mass energies, together with the FCC-hh, will provide the necessary complementary probes, enabling a far more complete exploration of the SMEFT parameter space.
\end{abstract}

\maketitle


\section{\label{sec:intro} Introduction}
The next generation of circular electron-positron colliders, such as the proposed Future Circular Collider (FCC-ee) and the Circular Electron-Positron Collider (CEPC), have the capability to operate at the $Z$-boson threshold, producing an unprecedented number of $Z$ bosons. This high-statistics environment enables precise measurements of electroweak precision observables (EWPOs), allowing for the high-sensitivity testing of the Standard Model (SM). Due to the high luminosity of these experiments, TeraZ, referring to the production of approximately $10^{12}$ $Z$-bosons, offers a window into potential deviations from the SM, thereby serving as a crucial indirect probe of new physics.

Recent studies have highlighted the power of EWPO measurements in setting stringent and model-independent constraints on physics beyond the SM (BSM) \cite{Gargalionis:2024jaw, Maura:2024zxz, Allwicher:2024sso}. It is argued that the precision nature of these observables makes them an essential tool in the search for new physics, as any deviation from SM predictions could signal the presence of new interactions or particles. Historically, a scientific outcome of the Large Electron-Positron Collider (LEP) was that such measurements provide sensitive probes to {\it some} BSM physics scenarios, guiding theoretical developments and placing constraints on new physics models\footnote{A well-known counterexample to the claimed sensitivity of LEP-measured EWPOs to new physics is the scenario of a sequential fourth generation of fermions. In this case, the presence of fourth-generation fermions with masses of a few hundred GeV ---i.e., not far above the electron-positron collision energy--- could not be excluded due to a blind (or insensitive) direction in the $S-T$ parameter plane \cite{Kribs:2007nz}. Thus, only Higgs boson measurements excluded a fourth generation at the LHC.}. 

Nowadays, the standard approach to interpreting small deviations from the SM is the framework of Effective Field Theory (EFT), which provides a systematic method for parametrising BSM effects in a model-independent manner. The power of indirect searches has been well established in flavour physics, where the Low-Energy Effective Field Theory (LEFT) framework has been instrumental in placing strong constraints on BSM scenarios~\cite{Buchalla:1995vs,Buras:2000dm,Buttazzo:2017ixm}. However, when moving to the SMEFT, the complexity of the problem increases significantly. Unlike LEFT, which operates at lower energies with a relatively small operator basis, SMEFT introduces a vastly larger operator basis, comprising 2,499 independent dimension-six operators~\cite{Grzadkowski:2010es}, many of which contribute to the 26 measured EWPOs. Moreover, this high-dimensional parameter space introduces the possibility of cancellations, where multiple SMEFT operators can conspire to mask new physics contributions, leading to potential blind spots in indirect searches.\footnote{High-energy hadron colliders possess the advantage to be able to break such blind directions by exploiting different kinematic regions of the phase space, where the operators with non-trivial energy dependence contribute overproportionally compared to the SM, therefore being constrained more tightly \cite{Englert:2015hrx,Englert:2017aqb,Franceschini:2017xkh,Banerjee:2018bio,Belvedere:2024wzg,Davighi:2024syj}.}

Given the many model parameters and limited physical observables, a key question arises: {\it Could new physics be just around the corner yet remain undetectable at TeraZ due to such operator cancellations or blind directions?}

Addressing this question requires a deeper understanding of how concrete ultraviolet (UV) models project onto the SMEFT parameter space. 
To the best of our knowledge, this question has not yet been explored in a fully comprehensive manner. While it has been recognized that a fully agnostic variation of EFT operator contributions to EWPOs gives rise to blind directions~\cite{Maura:2024zxz}, their relevance in more realistic UV scenarios has received comparatively little attention. In many cases, only a single field has been integrated out, leading to a restricted set of tightly correlated EFT operators contributing to the EWPOs, and thus avoiding blind directions.~\cite{DasBakshi:2020pbf,Maura:2024zxz,Gargalionis:2024jaw,Allwicher:2024sso}. 
Consequently, considering only such simplified scenarios may not fully capture the reach of the TeraZ programme, since realistic UV models typically involve several new degrees of freedom. In particular, multi-field extensions of the SM can generate correlations among SMEFT coefficients that give rise to situations where new physics lies at scales close to those probed by the HL-LHC, yet remains difficult to access at TeraZ. 
In this work, we make a first investigation of these correlations by exploring simple extensions of the SM and assessing whether they can evade detection at TeraZ. 

A fully comprehensive and exhaustive treatment of all technical aspects is beyond the scope of this work. We introduce controlled simplifications in specific technical steps to maintain clarity in our argument. 
It is worth noting, though, that relaxing any of these assumptions would generally worsen the issue of blind directions in EWPO constraints.
First, in our analysis, we refrain from marginalising over all effective operators
since otherwise it could obscure the underlying correlations that relate 
blind spots to the dynamics of UV model realisations. Second, we restrict to the sub-sector of the SMEFT spanned by four-fermion interactions with only third-family couplings, and we avoid exploring blind directions that involve unreasonably different sizes of Wilson coefficients (WCs).

Starting with LEP data first, in Section~\ref{sec:bottomup} we display the resulting blind directions of the so-called bottom-up approach, i.e. agnostically varying EFT operators contributing to the EWPOs. Then, in Section~\ref{sec:uvspace}, we consider which UV completions of the SMEFT generate the different blind directions upon integrating the new heavy fields out at tree level. In Section~\ref{sec:multi}, we discuss the fit of selected scenarios to EWPO, including one-loop matching corrections. In Section~\ref{sec:terazcase}, we examine how these considerations evolve at TeraZ.
We offer a summary and conclusions in Section~\ref{sec:summary}.

\section{\label{sec:bottomup} Effective Field Theory Fit: bottom up}
\begin{figure}[ht]
 \centering
 \includegraphics[width=0.49\columnwidth]{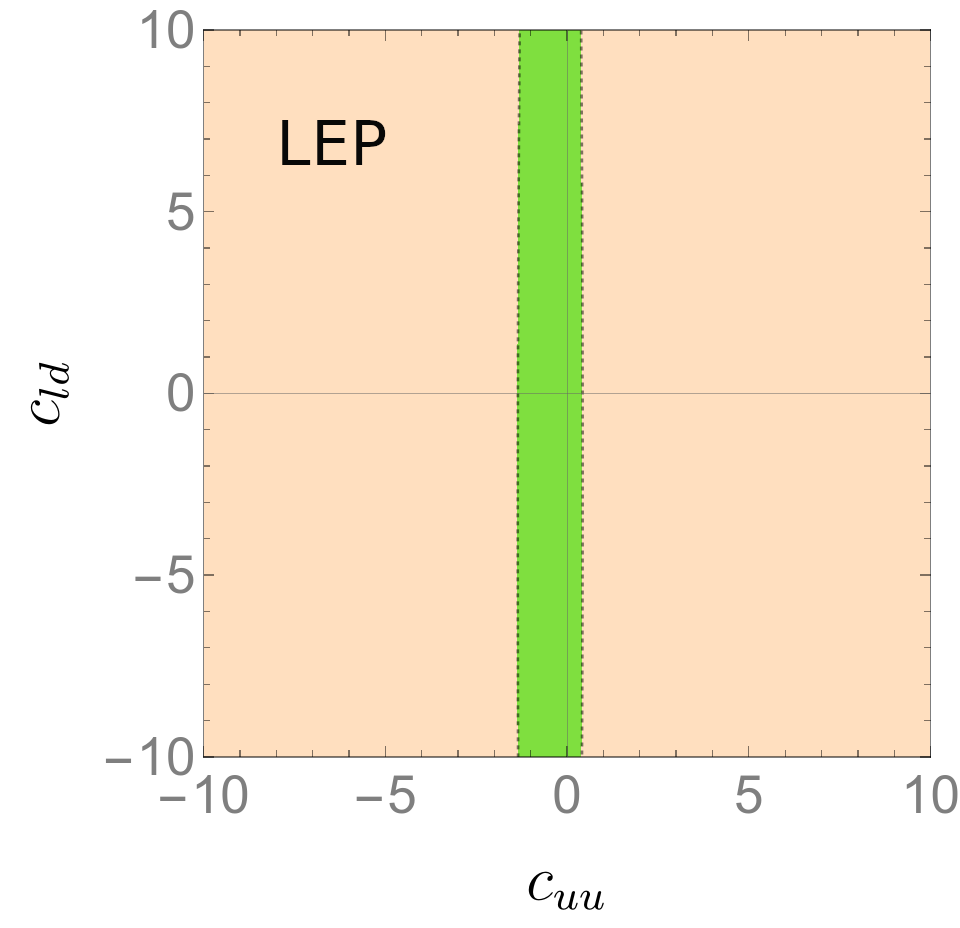}
 \includegraphics[width=0.49\columnwidth]{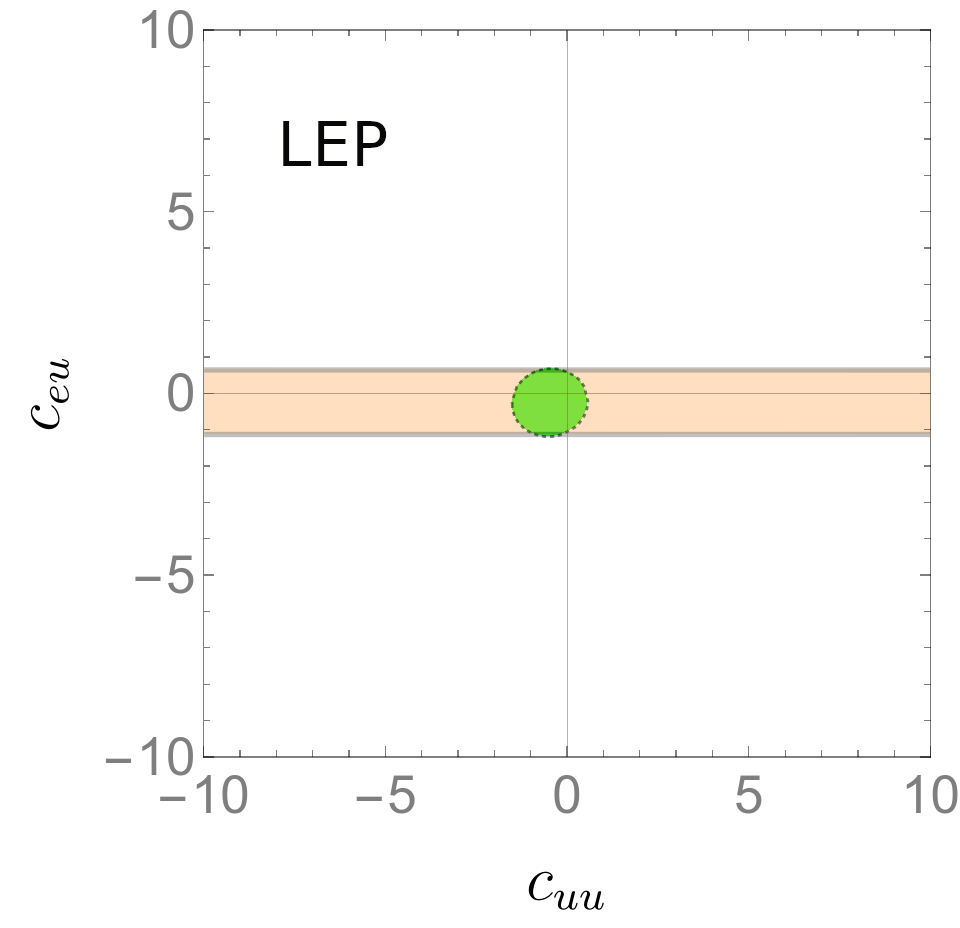}
 \caption{\it Examples of EWPO bounds at 3\,$\sigma$ on four-fermion regions involving $c_{uu}$. The orange area enclosed by a solid line includes leading-log RGEs, while we integrate RGEs numerically in the green area enclosed by the dashed line.}\label{fig:uu}
\label{fig:ewpd4f}
\end{figure}

In this section, we adopt a bottom-up EFT approach to explore potential blind directions in EWPOs from LEP, assuming minimal prior knowledge about the UV origin of the effective operators. We also assess the impact of renormalisation group evolution (RGE), incorporating leading-log RGEs and full numerical integration of the one-loop SMEFT RGEs~\cite{Jenkins:2013wua,Jenkins:2013zja,Alonso:2013hga}.

We consider a total of 26 EWPO:
\begin{align}
    \text{EWPO} = \bigg\lbrace &\Gamma_W, \Gamma_W^{e\nu,\mu\nu,\tau\nu},\Gamma^\text{had}_W, \sigma_\text{had}, \Gamma_Z, A_\text{FB}^{e,\mu,\tau},\nonumber\\
    &A_\text{FB}^{s,c,b}, A_{e,\mu,\tau}, A_{s,c,b}, R_{e,\mu,\tau}, R_{s,c,b}, \alpha\bigg\rbrace\,,
\end{align}
which represent the total decay width of the $W$, the partial width of the $W$ into different leptons, the partial width of the $W$ into quarks, the total $Z$ hadronic cross section, the $Z$ total decay rate, the forward-backward and left-right asymmetries as well as the ratios of the $Z$ decays into different quarks and leptons and the fine-structure constant. The definitions of these parameters and experimental values can be found in Refs.~\cite{ATLAS:2014wva,Davier:2010nc,Group:2012gb,ALEPH:2005ab,ALEPH:2010aa,ParticleDataGroup:2024cfk,Biekotter:2025nln}. We take the Fermi constant $G_F$ and the $W$ mass $m_W$ as input parameters.

When restricting to CP-conserving, baryon-number-preserving operators in the Warsaw basis~\cite{Grzadkowski:2010es} and neglecting flavour structure, we find $\mathcal{O}(50)$ operators, of which 10 independent operators can contribute to EWPOs at tree level~\cite{deBlas:2015aea}. Including RGE evolution, the number of operators contributing to EWPOs grows to 31, out of which 25 can arise from integrating out heavy fields at tree level in realistic UV completions of the SM. While it seems that the 26 EWPOs could overconstrain such a system, not all operators contribute to all EWPOs, and, thus, even imposing such strong constraints on the operators, one finds rather unconstrained subspaces of operators that lead to blind directions. For the numerical results of the fits, we use current constraints on the EWPOs, as implemented in \texttt{smelli}~\cite{Aebischer:2018iyb,Stangl:2020lbh}.

We focus on the space spanned by four-fermion operators,\footnote{Note that blind directions within the rest of the SMEFT include the obvious cases of $\mathcal{O}_\phi$ ---which affects only the Higgs potential~\cite{deBlas:2014mba}---, $\mathcal{O}_{\phi\square}$ ---which enters weakly via RGEs--- as well as certain operators that arise only at loop level. Note also that the combination $c_{\phi q}^{(1)}-c_{\phi q}^{(3)}$ is untestable with EWPO but only if it holds right at the $Z$ pole; RGE removes this one.} as they are ubiquitous in models of new physics, arise easily at tree level in UV completions of the SMEFT and, with the exception of $\mathcal{O}_{ll}$ with first family leptons, none enters into EWPO at tree level~\cite{deBlas:2015aea}. The most unconstrained four-fermions involve third-family fermions only, in which we concentrate. Theoretically, one can motivate that new physics couples mostly to the third generation~\cite{Contino:2003ve,Panico:2015jxa,Davighi:2023iks,Chung:2023gcm}.  We also ignore $c_{quqd}^{(1)}$, $c_{quqd}^{(8)}$, $c_{ledq}$, $c_{lequ}^{(1)}$ and $c_{lequ}^{(3)}$, as they are in general flavour-violating. We follow the notation of Ref.~\cite{Grzadkowski:2010es} for SMEFT WCs.

We assume WCs to be defined at a scale $\Lambda=1$ TeV, and study their impact on EWPO at the scale $\sim m_Z$ using the leading-log dimension-six SMEFT RGEs, as well as upon integrating numerically these RGEs. To this latter goal, we rely on \texttt{smelli} which is further based on \texttt{flavio}~\cite{Straub:2018kue} and \texttt{Wilson}~\cite{Aebischer:2018bkb}.

To start with, the following WCs do not contribute to EWPO or their contribution is negligible: $c_{uu}$, $c_{dd}$, $c_{qu}^{(8)}$, $c_{qd}^{(8)}$, $c_{ud}^{(8)}$, $c_{le}$, $c_{ll}$, $c_{ed}$, $c_{ld}$. This explains why most of these WCs do not appear in Fig.~1 of Ref.~\cite{Maura:2024zxz}. Out of these, though, $c_{uu}$ is known to be well constrained by two-loop effects captured by RGE re-summation~\cite{Allwicher:2023aql,Allwicher:2023shc,Stefanek:2024kds,Maura:2024zxz}; see Fig.~\ref{fig:uu}.

\begin{figure}
 \centering
 \includegraphics[width=0.49\columnwidth]{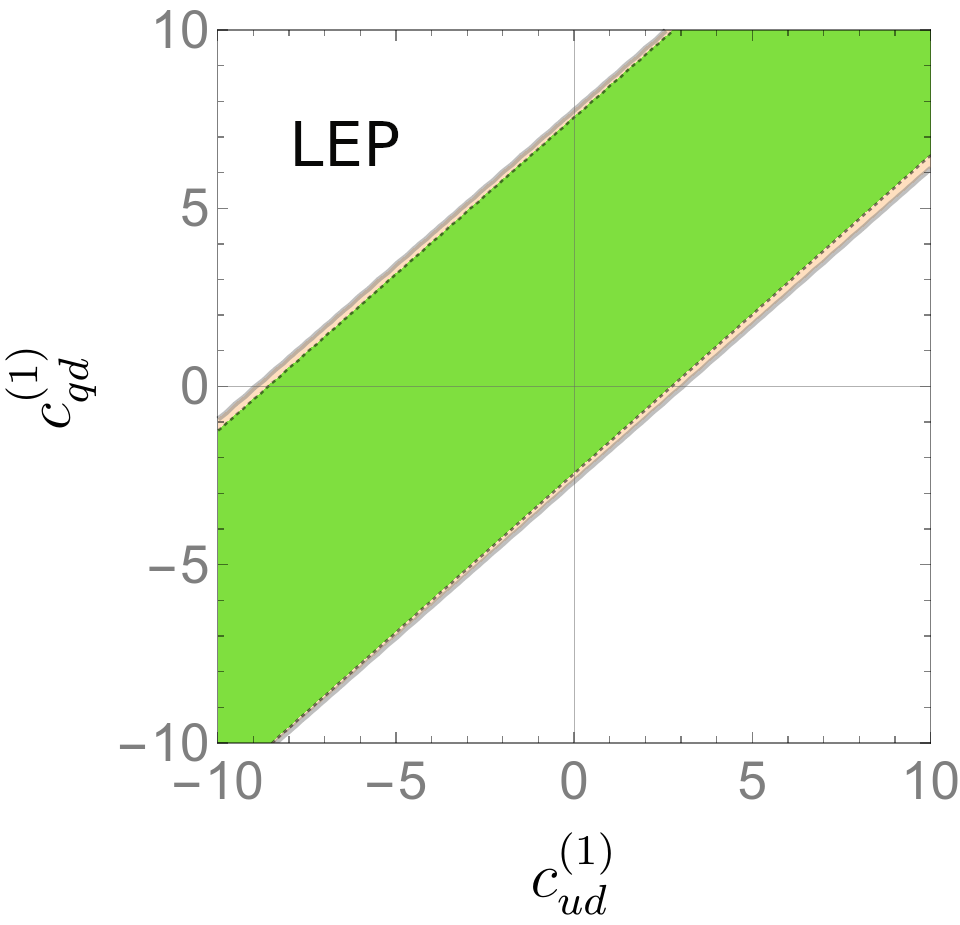}
 \includegraphics[width=0.49\columnwidth]{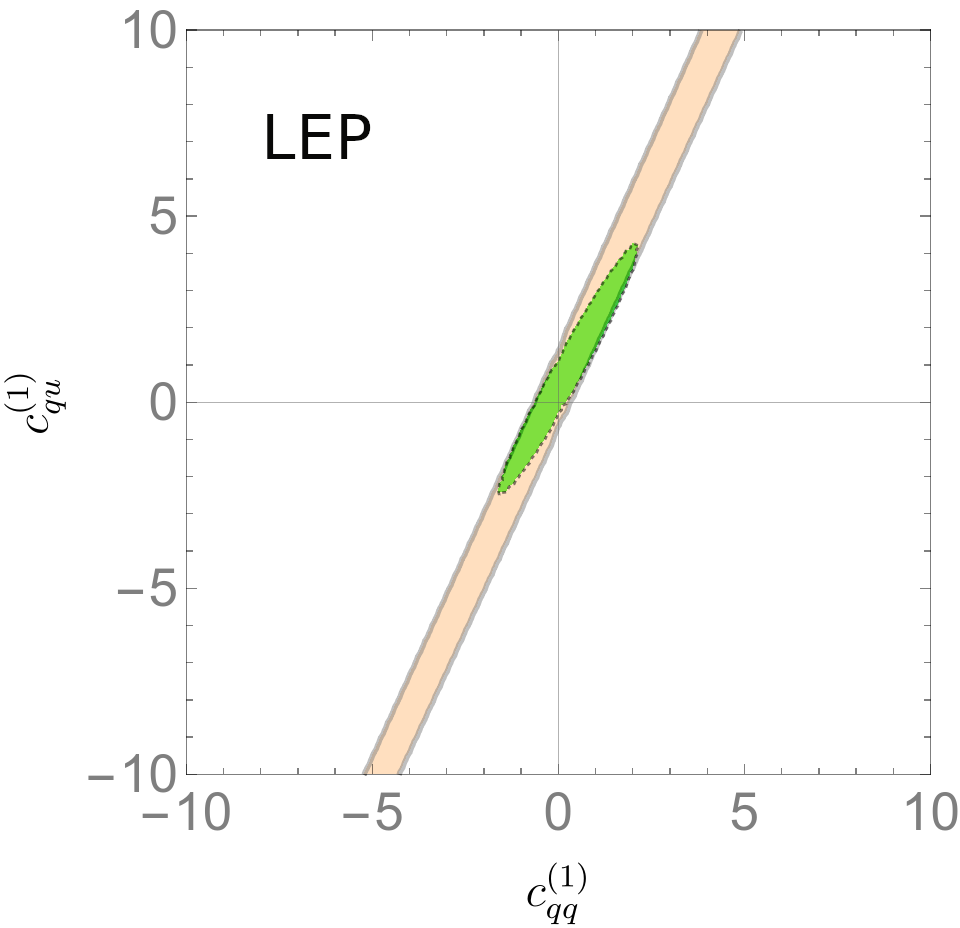}
 \includegraphics[width=0.49\columnwidth]{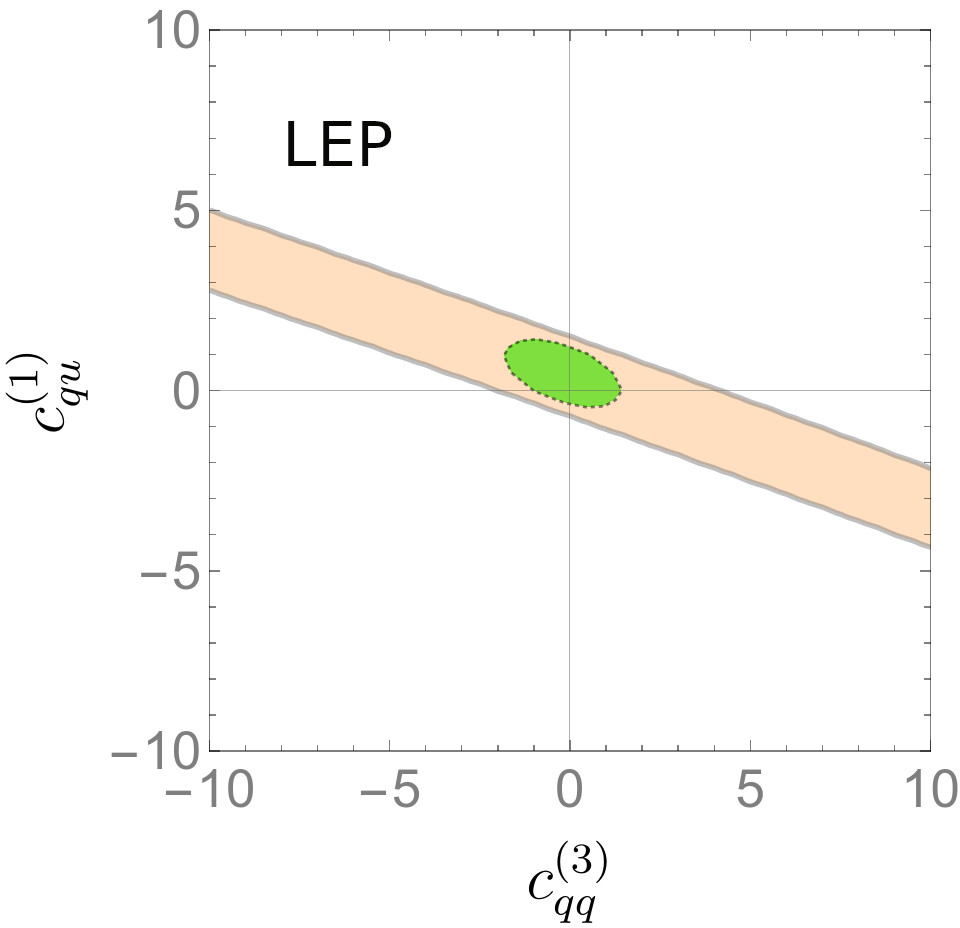}
 \includegraphics[width=0.49\columnwidth]{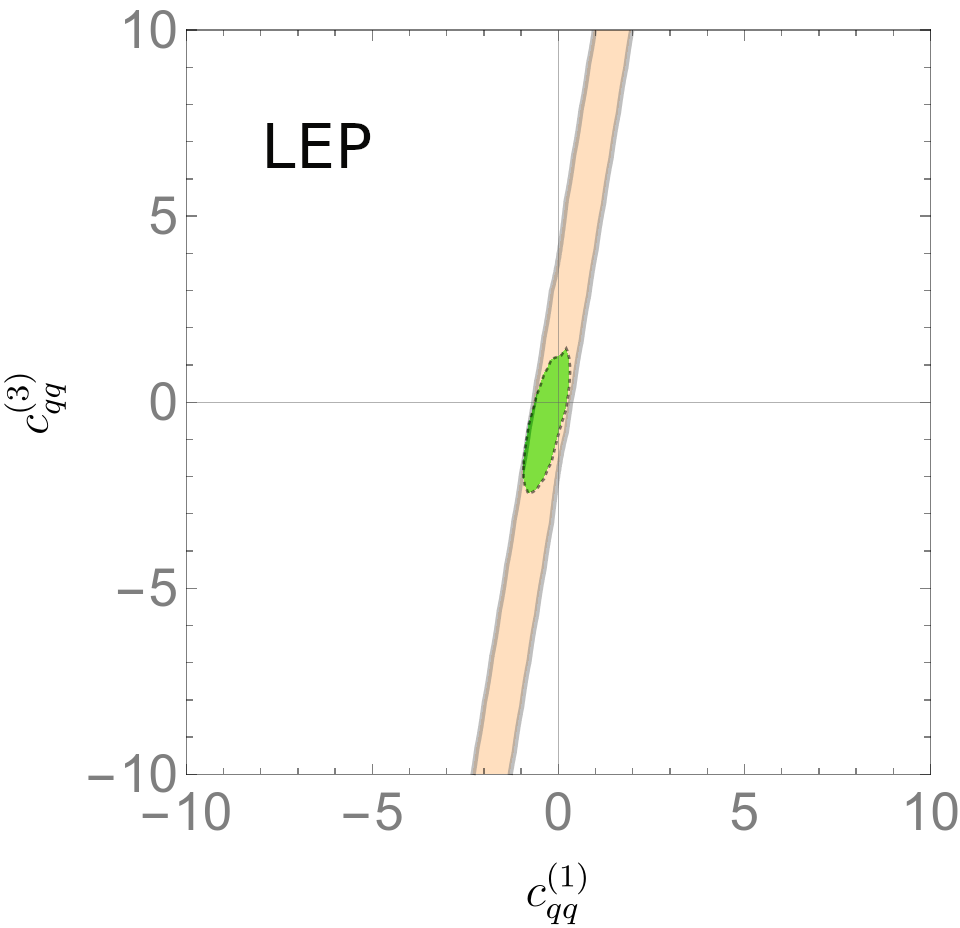}
 \caption{\it Regions in the space of $c_{ud}^{(1)}$, $c_{qd}^{(1)}$, $c_{qq}^{(1)}$, $c_{qu}^{(1)}$ and $c_{qq}^{(3)}$ excluded by EWPOs at  $3\,\sigma$. The orange region enclosed by the solid line includes only leading-log RGE effects, while we numerically integrate the RGEs to obtain the green region enclosed by the dashed line.}\label{fig:blinddirections}
\end{figure}

On top of this, we explore all sub-spaces spanned by pairs of the Warsaw-basis four-fermion operators in the search for blind directions. Assuming no disparate sizes for ratios of WCs, we find blind directions in the sub-spaces spanned by $\lbrace c_{qd}^{(1)}, c_{ud}^{(1)}\rbrace$, $\lbrace c_{qq}^{(1)},c_{qu}^{(1)} \rbrace$, $\lbrace c_{qq}^{(3)},c_{qu}^{(1)}\rbrace$ and $\lbrace c_{qq}^{(1)}, c_{qq}^{(3)} \rbrace$. However, the last three are gone when RGE resummation is considered;\footnote{Approximate flat directions appear also in the planes of $\{c_{eu}, c_{qe}\}$ and $\{c_{lu}, c_{lq}^{(1)}\}$, but here we focus on most robust ones involving all them.} see Fig.~\ref{fig:blinddirections}. For the one that survives, we have, in TeV: 
\begin{align}
\begin{pmatrix}\Gamma_Z\\A_b\\R_{e,\mu,\tau}\end{pmatrix} \sim \begin{pmatrix} -0.0009 & 0.0008\\ -0.004 & 0.004 \\ -0.01 & 0.009  \end{pmatrix} \begin{pmatrix}c_{ud}^{(1)}\\c_{qd}^{(1)}\end{pmatrix}\,,
\end{align}
where we have included the EWPOs that receive the largest contributions from these WCs. The null vector of the corresponding matrix, which describes the blind direction in this sub-space, is given by: $c_{ud}^{(1)}\sim c_{qd}^{(1)}$. It extends to the full set of EWPOs. 

\begin{figure}[t]
 \centering
 \includegraphics[width=0.49\columnwidth]{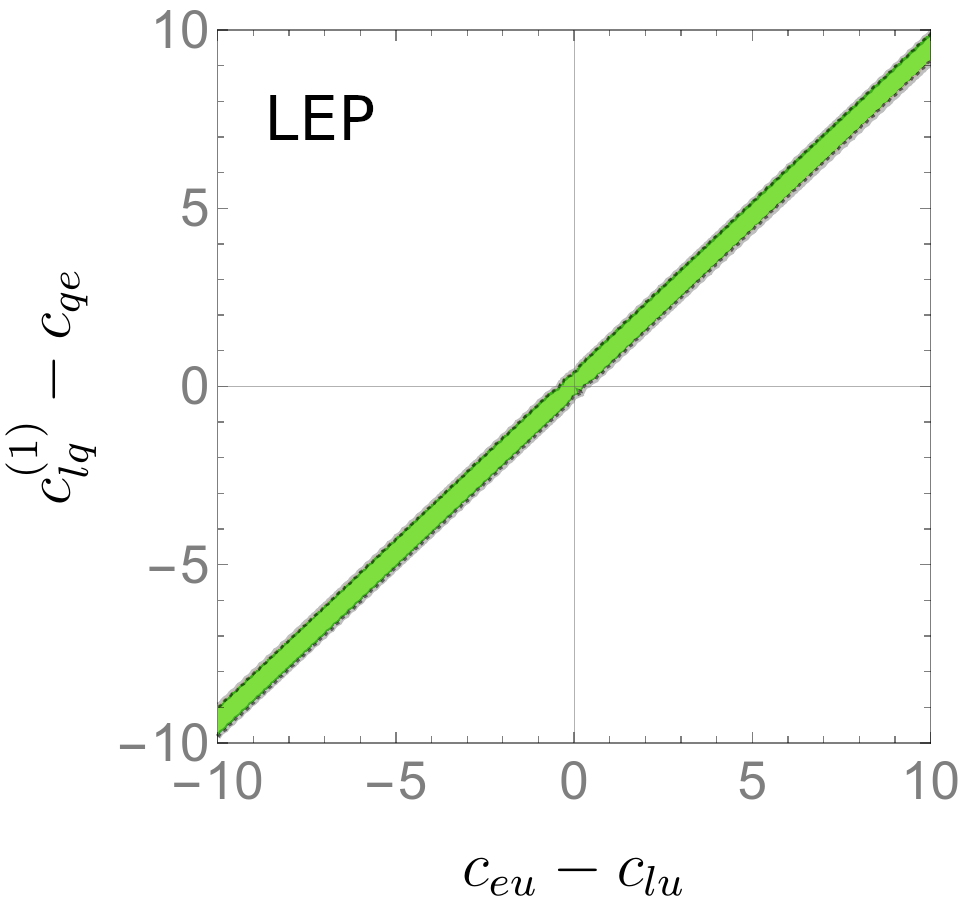}
 \includegraphics[width=0.49\columnwidth]{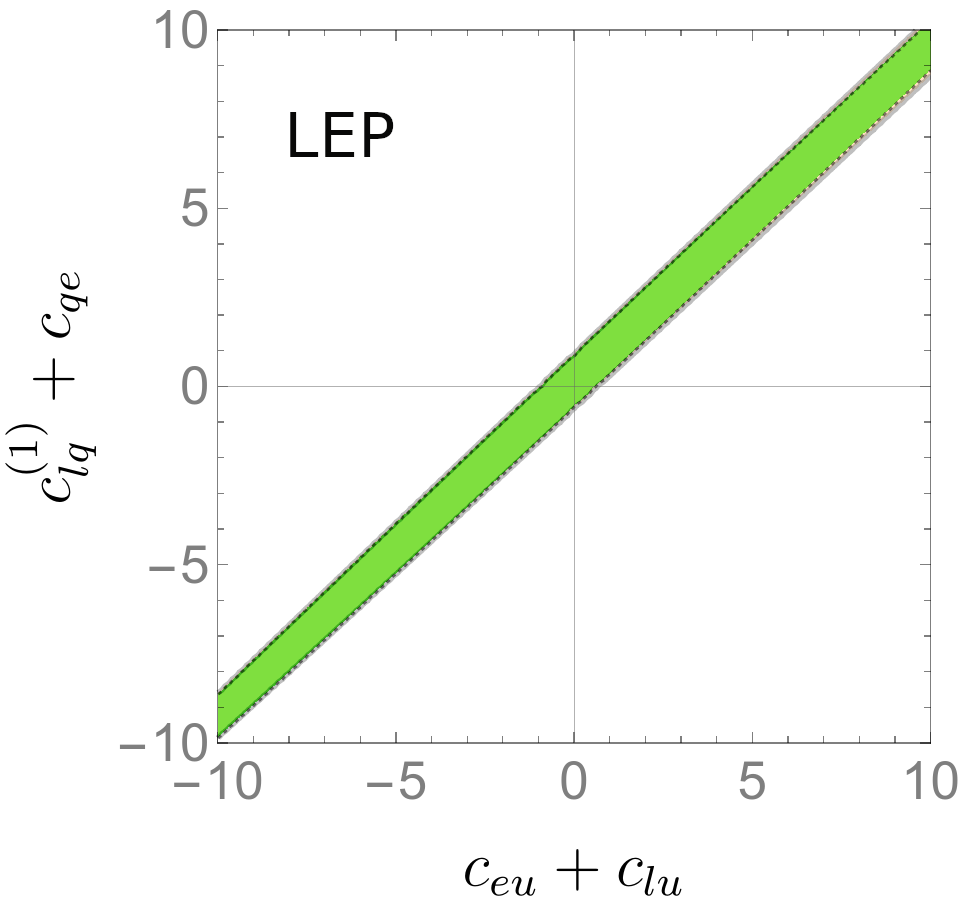}
 \caption{\it Regions in the space of $c_{eu}$, $c_{lu}$, $c_{qe}$ and $c_{lq}^{(1)}$ excluded by EWPO at $3\,\sigma$. The orange region enclosed by the solid line includes only leading-log RGE effects, while we numerically integrate the RGEs to obtain the green one enclosed by the dashed line.}\label{fig:higherblinddirections}
\end{figure}

There are no blind directions involving three four-fermions (assuming no disparate coefficients as well as survival under RGE resummation). With four, however, we find $c_{eu}\sim c_{qe}\sim c_{lu}\sim c_{lq}^{(1)}$ as well as  $c_{eu}\sim c_{qe}\sim -c_{lu}\sim -c_{lq}^{(1)}$; see Fig.~\ref{fig:higherblinddirections}. Indeed, we have, in TeV:
\begin{align}\label{eq:second_blinddirection}
    \begin{pmatrix}A_\tau\\ R_\tau\end{pmatrix} \sim \begin{pmatrix}-0.01 & 0.01 & -0.01 & 0.01\\-0.19 & -0.24 & 0.23 & 0.20\end{pmatrix}\begin{pmatrix}c_{eu}\\ c_{lq}^{(1)} \\ c_{lu} \\ c_{qe}\end{pmatrix}\,.
\end{align}

\begin{table*}
\centering
\resizebox{!}{5.5cm}{%
\begin{tabular}{lll}
  \toprule
  \multicolumn{2}{l}{Field}
  & Effective Lagrangian \\
  \midrule
  $\Xi_1$ & $(1, 3)_1$
  &
    $+ \frac{|y_{\Xi_1}|^2}{M_{\Xi_1}^{2}} \mathcal{O}_{ll}$
  \\[7pt]
  $\varphi$ & $(1, 2)_{1/2}$
  &
    $- \frac{|y^e_{\varphi}|^2}{ 2 M_{\varphi}^{2}} \mathcal{O}_{le}
    - \frac{|y^d_{\varphi}|^2}{6 M_{\varphi}^{2}}
    \left(\mathcal{O}^{(1)}_{qd} + 6 \mathcal{O}^{(8)}_{qd}\right)$
  \\[7pt]
  \multirow{2}{*}{$\omega_1$} & 
  \multirow{2}{*}{$(3, 1)_{-1/3}$}
  &
    $+ \frac{|y^{du}_{\omega_1}|^2}{3 M_{\omega_1}^2}
    \left(\mathcal{O}^{(1)}_{ud} - 3 \mathcal{O}^{(8)}_{ud}\right)
     + \frac{|y^{eu}_{\omega_1}|^2}{ 2 M_{\omega_1}^{2}} \mathcal{O}_{eu}$
  \\
  & &
     $+ \frac{|y^{ql}_{\omega_1}|^2}{ 4 M_{\omega_1}^{2}} 
       \left(\mathcal{O}^{(1)}_{lq} - \mathcal{O}^{(3)}_{lq}\right)$
  \\[7pt]
  $\omega_4$ & $(3, 1)_{-4/3}$
  &
    $+ \frac{|y^{ed}_{\omega_4}|^2}{ 2 M_{\omega_4}^{2}
    \mathcal{O}_{ed}}$
  \\[7pt]
  $\Pi_1$ & $(3, 2)_{1/6}$
  &
    $- \frac{|y_{\Pi_1}|^2}{ 2 M_{\Pi_1}^2}  \mathcal{O}_{ld}$
  \\[7pt]
  $\Pi_7$ & $(3, 2)_{7/6}$
  &
    $- \frac{|y^{eq}_{\Pi_7}|^2}{ 2 M_{\Pi_7}^{2}} \mathcal{O}_{qe}
     - \frac{|y^{lu}_{\Pi_7}|^2}{ 2 M_{\Pi_7}^{2}} \mathcal{O}_{lu}$
  \\[7pt]
  $\zeta$ & $(3, 3)_{-1/3}$
  &
    $+ \frac{|y^{ql}_{\zeta}|^2}{4 M_{\zeta}^{2}} 
       \left(3 \mathcal{O}^{(1)}_{lq} + \mathcal{O}^{(3)}_{lq}\right)$
  \\[7pt]
  $\Omega_1$ & $(6, 1)_{1/3}$
  &
    $+ \frac{|y^{ud}_{\Omega_1}|^2}{6 M_{\Omega_1}^{2}}
    \left(2 \mathcal{O}^{(1)}_{ud} + 3 \mathcal{O}^{(8)}_{ud}\right)$
  \\[7pt]
  $\Omega_2$ & $(6, 1)_{-2/3}$
  &
    $+ \frac{|y_{\Omega_2}|^2}{ 2 M_{\Omega_2}^{2}}  \mathcal{O}_{dd}$
  \\[7pt]
  $\Phi$ & $(8, 2)_{1/2}$
  &
    $- \frac{|y^{dq}_{\Phi}|^2}{18 M_{\Phi}^{2}}
    \left(4 \mathcal{O}^{(1)}_{qd} - 3 \mathcal{O}^{(8 )}_{qd}\right)$
  \\[7pt]
  \bottomrule
\end{tabular}
}
\qquad
\resizebox{!}{5.5cm}{%
\begin{tabular}{lll}
  \toprule
  \multicolumn{2}{l}{Field}
  & Effective Lagrangian \\
  \midrule
  $\mathcal{B}$ & $(1, 1)_0$
  &
    $- \frac{|g^d_{\mathcal{B}}|^2}{ 2 M_{\mathcal{B}}^{2}}  \mathcal{O}_{dd}
    \,
    - \frac{|g^l_{\mathcal{B}}|^2}{2 M_{\mathcal{B}}^{2}}
    \mathcal{O}_{ll}$
  \\[7pt]
  $\mathcal{B}_{1}$ & $(1, 1)_{1}$
  &
    $- \frac{|g^{du}_{\mathcal{B}_1}|^2}{3 M_{\mathcal{B}_1}^{2}}
    \left(\mathcal{O}^{(1)}_{ud} + 6 \mathcal{O}^{(8)}_{ud}\right)$
  \\[7pt]
  $\mathcal{W}$ & $(1, 3)_0$
  &
    $- \frac{|g^l_{\mathcal{W}}|^2}{8 M_{\mathcal{W}}^2} \mathcal{O}_{ll}$
  \\[7pt]
  $\mathcal{G}$ & $(8, 1)_0$
  &
    $- \frac{|g^d_{\mathcal{G}}|^2}{4 M_{\mathcal{G}}^2}
    \mathcal{O}_{dd}$
  \\[7pt]
  $\mathcal{G}_1$ & $(8, 1)_1$
  &
    $- \frac{|g_{\mathcal{G}_1}|^2}{9 M_{\mathcal{G}_1}^{2}}
    \left(4 \mathcal{O}^{(1)}_{ud} - 3 \mathcal{O}^{(8)}_{ud}\right)$
  \\[7pt]
  $\mathcal{L}_3$ & $(1, 2)_{-3/2}$
  &
    $+ \frac{|g_{\mathcal{L}_3}|^2}{  M_{\mathcal{L}_3}^2}
    \mathcal{O}_{le}$
  \\[7pt]
  $\mathcal{U}_2$ & $(3, 1)_{2/3}$
  &
    $- \frac{|g^{ed}|^2}{  M_{\mathcal{U}_2}^{2}}  \mathcal{O}_{ed}
    - \frac{|g^{lq}_{\mathcal{U}_2}|^2}{2 M_{\mathcal{U}_2}^{2}}
      \left(\mathcal{O}^{(1)}_{lq} + \mathcal{O}^{(3)}_{lq}\right)$
  \\[7pt]
  $\mathcal{U}_5$ & $(3, 1)_{5/3}$
  &
    $- \frac{|g_{\mathcal{U}_5}|^2}{M_{\mathcal{U}_5}^{2}}
       \mathcal{O}_{eu}$
  \\[7pt]
  $\mathcal{Q}_1$ & $(3, 2)_{1/6}$
  &
    $+ \frac{2 |g^{dq}_{\mathcal{Q}_1}|^2}{3 M_{\mathcal{Q}_1}^{2}}
    \left(\mathcal{O}^{(1)}_{qd} - 3 \mathcal{O}^{(8)}_{qd}\right)
    + \frac{|g^{ul}_{\mathcal{Q}_1}|^2}{M_{\mathcal{Q}_1}^{2}}
       \mathcal{O}_{lu}$
  \\[7pt]
  $\mathcal{Q}_5$ & $(3, 2)_{-5/6}$
  &
    $+ \frac{|g^{dl}_{\mathcal{Q}_5}|^2}{  M_{\mathcal{Q}_5}^2}
    \mathcal{O}_{ld}
     + \frac{|g^{eq}_{\mathcal{Q}_5}|^2}{M_{\mathcal{Q}_5}^{2}}
       \mathcal{O}_{qe}$
  \\[7pt]
  $\mathcal{X}$ & $(3, 3)_{2/3}$
  &       
    $- \frac{|g_{\mathcal{X}}|^2}{8 M_{\mathcal{X}}^{2}}
       \left(3 \mathcal{O}^{(1)}_{lq} - \mathcal{O}^{(3)}_{lq}\right)$
  \\[7pt]
  $\mathcal{Y}_1$ & $(\bar{6}, 2)_{1/6}$
  &       
    $+ \frac{|g_{\mathcal{Y}_1}|^2}{3 M_{\mathcal{Y}_5}^{2}}
    \left(2 \mathcal{O}^{(1)}_{qd} + 3 \mathcal{O}^{(8)}_{qd}\right)$
  \\
  \bottomrule
\end{tabular}
}
\label{tab:matching}
\caption{\it Tree-level matching contributions to potential blind directions from scalars (left) and vectors (right).
It is assumed that there is only one non-vanishing coupling to the SM per heavy field, so only one of the terms shown in the expression for the effective Lagrangian generated by each field is present at the same time.}
\end{table*}

We refrain from considering larger combinations of WCs for the sake of clarity and tractability. Moreover, as demonstrated in the following section, the class of UV completions that naturally populate the blind subspaces identified above is sufficiently broad to substantiate our main conclusions.

Altogether, we focus on the following combinations of four-fermion operators:
\begin{align}\label{eq:blinddirections}
    \lbrace &\mathcal{O}_{dd}, \mathcal{O}_{qu}^{(8)},
    \mathcal{O}_{qd}^{(8)}, \mathcal{O}_{ud}^{(8)},
    \mathcal{O}_{le},
    \mathcal{O}_{ll}, 
    \mathcal{O}_{ed}, \mathcal{O}_{ld}, 
    \nonumber \\
    &\mathcal{O}_{ud}^{(1)}+\mathcal{O}_{qd}^{(1)},
    \nonumber\\
    &\mathcal{O}_{eu}+\mathcal{O}_{qe}+\mathcal{O}_{lu}
    +\mathcal{O}_{lq}^{(1)},
    \nonumber \\
    &\mathcal{O}_{eu}+\mathcal{O}_{qe}-\mathcal{O}_{lu}
    -\mathcal{O}_{lq}^{(1)}\rbrace\,.
\end{align}
We emphasize that this is not a complete list of all blind directions, but just the simplest ones we have found with the stated assumptions. As we will see in the next sections, this is sufficient to illustrate 
potential blind spots at TeraZ, since there are several extensions of the SM that generate each of them.

We further observe that finite one-loop SMEFT effects~\cite{Biekotter:2025nln} slightly modify, but do not eliminate, the blind directions identified. In the following section, we systematically examine extensions of the SM that, when matched at tree level onto the SMEFT, give rise to effective operators aligned with these blind subspaces.

\section{\label{sec:uvspace} UV extensions of the SMEFT}

The complete classification of SM extensions that generate dimension-six SMEFT operators at tree level was presented in Ref.~\cite{deBlas:2017xtg}, building on earlier work~\cite{delAguila:2000rc,delAguila:2008pw,deBlas:2014mba}. For clarity, we reproduce in Table~\ref{tab:matching} only the subset relevant to our analysis, along with the corresponding effective operators they induce. We adopt the coupling notation used in Ref.~\cite{deBlas:2017xtg}, and note that when multiple heavy fields are present, their contributions to the SMEFT Lagrangian simply add at leading order.

We assume that each heavy field $X$ has a single coupling to the SM, schematically of the form $X \psi_1 \psi_2$, where $\psi_1$ and $\psi_2$ are SM fermions.
The contribution of such a field $X$ to the SMEFT Lagrangian at tree level and dimension six is a linear combination of one or two Warsaw-basis operators with the field content $(\psi_1)^2 (\psi_2)^2$.
The only pairs of operators that may be generated are 
$\{\mathcal{O}_{x}^{(1)}, \mathcal{O}_{x}^{(8)}\}$,
with $x = qu$, $qd$ and $ud$, and $\{\mathcal{O}_{lq}^{(1)}, \mathcal{O}_{lq}^{(3)}\}$.
The effects of the $\mathcal{O}_x^{(8)}$ operators on EWPO are all negligible.
Thus, each heavy field we consider effectively generates a single operator relevant for EWPO, except for the ones that generate the $\mathcal{O}_{lq}$ operators. For the latter, we always consider heavy fields in pairs in which $c_{lq}^{(3)}$ is cancelled, so again only one operator (i.e. $\mathcal{O}_{lq}^{(1)}$) survives.

Our aim is to identify which tree-level completions of the SMEFT, involving a limited number of heavy fields, give rise to effective interactions aligned with the blind subspace defined in Eq.~\eqref{eq:blinddirections}.
We make use of \texttt{MatchingDB}~\cite{Aebischer:2023nnv} for this purpose.

We start with the following single-field extensions that generate only an unconstrained operator:
\begin{itemize}
    \item $\Xi_1$, $\mathcal{B}$ and $\mathcal{W}$ generate $\mathcal{O}_{ll}$.
    \item $\varphi$ and $\mathcal{L}_3$ generate $\mathcal{O}_{le}$.
    \item 
    $\Omega_2$, $\mathcal{B}$ and $\mathcal{G}$ generate $\mathcal{O}_{dd}$.
    \item $\omega_4$ and $\mathcal{U}_2$ generate $\mathcal{O}_{ed}$.
    \item $\Pi_1$ and $\mathcal{L}_3$ generate $\mathcal{O}_{ld}$.
\end{itemize}

The blind direction $c^{(1)}_{ud} \simeq c^{(1)}_{qd}$ can not be generated by single-field extensions; at least two fields are needed, and it can be read from Table~\ref{tab:matching}.
It involves $\varphi$ or $\Phi$ with either $\mathcal{B}_1$ or $\mathcal{G}_1$ as well as $\omega_1$ or $\Omega_1$ with either $\mathcal{Q}_1$ or $\mathcal{Y}_1$.

Concerning $c_{eu} \sim c_{qe} \sim \pm c_{lu} \sim \pm c^{(1)}_{lq}$, there are eight five-field extensions that generate this blind direction.
They are given by any combination of a pair of fields from the list
\begin{equation}
    \{(\omega_1, \mathcal{Q}_5), (\mathcal{U}_5, \Pi_7)\}
\end{equation}
with a triad of fields from the list
\begin{equation}
    \{(\Pi_7, \omega_1, \mathcal{X}),
      (\Pi_7, \mathcal{U}_2, \mathcal{X}),
      (\mathcal{Q}_1, \omega_1, \zeta),
      (\mathcal{Q}_1, \zeta, \mathcal{U}_2)\}\,.
\end{equation}
If any of the fields is repeated in the combination, it is understood that two copies of that field are present, each with different couplings to the SM. All these possible combinations are displayed in Table~\ref{tab:extensions}.

\begin{table}
\centering
\begin{tabular}{ccc}
  \toprule
  Blind direction && Extensions \\
  \midrule
  $\mathcal{O}_{ll}$ && $\{\Xi_1\} \quad \{\mathcal{B}\} \quad \{\mathcal{W}\}$ \\
  $\mathcal{O}_{le}$ && $\{\varphi\} \quad \{\mathcal{L}_3\}$ \\
  $\mathcal{O}_{dd}$ && $ 
  \{\Omega_2\} \quad \{\mathcal{G}\}$ \\
  $\mathcal{O}_{ed}$ && $\{\omega_4\} \quad \{\mathcal{U}_2\}$ \\
  $\mathcal{O}_{ld}$ && $\{\Pi_1\} \quad \{\mathcal{L}_3\}$ \\
  \midrule
  \multirow{2}{*}{$\mathcal{O}^{(1)}_{ud} + \mathcal{O}^{(1)}_{qd}$}
  && $\{\varphi, \mathcal{B}_1\} \quad \{\varphi, \mathcal{G}_1\} \quad
     \{\omega_1, \mathcal{Q}_1\} \quad \{\omega_1, \mathcal{Y}_1\}$ \\
  && $\{\Omega_1, \mathcal{Q}_1\} \quad \{\Omega_1, \mathcal{Y}_1\} \quad
    \{\Phi, \mathcal{B}_1\} \quad \{\Phi, \mathcal{G}_1\}$ \\
  \midrule
  $\mathcal{O}_{eu} + \mathcal{O}_{qe}$
  && $\{\omega_1, \mathcal{Q}_5, \mathcal{Q}_1, \omega_1, \zeta\} \quad
     \{\omega_1, \mathcal{Q}_5, \mathcal{Q}_1, \zeta, \mathcal{U}_2\}$ \\
  $\qquad + \, \mathcal{O}_{lu} + \mathcal{O}^{(1)}_{lq}$
  && $\{\mathcal{U}_5, \Pi_7, \Pi_7, \omega_1, \mathcal{X}\} \quad
    \{\mathcal{U}_5, \Pi_7, \Pi_7, \mathcal{U}_2, \mathcal{X}\}$ \\
  \midrule
  $\mathcal{O}_{eu} + \mathcal{O}_{qe}$
  && $\{\omega_1, \mathcal{Q}_5,  \Pi_7, \omega_1, \mathcal{X}\} \quad
     \{\omega_1, \mathcal{Q}_5, \Pi_7, \mathcal{U}_2, \mathcal{X}\}$ \\
  $\qquad - \, \mathcal{O}_{lu} - \mathcal{O}^{(1)}_{lq}$
  && $\{\mathcal{U}_5, \Pi_7, \mathcal{Q}_1, \omega_1, \zeta\} \quad
     \{\mathcal{U}_5, \Pi_7, \mathcal{Q}_1, \zeta, \mathcal{U}_2\}$ \\
  \bottomrule
\end{tabular}
\label{tab:extensions}
\caption{\it Extensions of the SM that generate blind directions.}
\end{table}

\begin{figure}
    \includegraphics[width=\columnwidth]{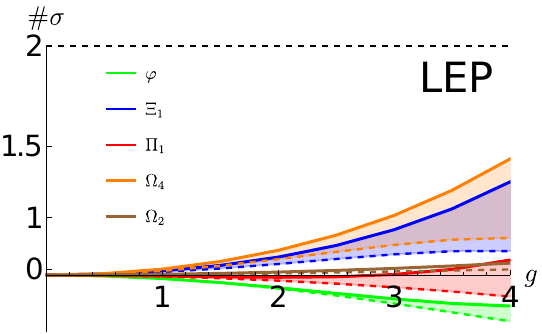}
    \caption{\it Sensitivity of EWPO to different single-scalar extensions of the SMEFT as a function of their only coupling to the SM. The solid (dash) line includes tree-level and leading-log RGE (RGE resummation and finite one-loop matching corrections).}\label{fig:scalars}
\end{figure}

One might expect that finite one-loop matching effects, those not already captured by renormalisation group evolution, could lift the blind directions. However, we have verified explicitly that this is not the case for the single-scalar extensions; see Fig.~\ref{fig:scalars}. To this end, we performed one-loop matching onto the SMEFT using dedicated routines, with the assistance of \texttt{SOLD}~\cite{Guedes:2023azv,Guedes:2024vuf} and \texttt{matchmakereft}~\cite{Carmona:2021xtq}. We note that integrating out massive vector fields at one loop is non-trivial without a complete description of the underlying spontaneously broken theory, and is currently not supported by existing automated tools~\cite{Carmona:2021xtq,Fuentes-Martin:2022jrf,Guedes:2023azv,Guedes:2024vuf}. For brevity and cohesion, we omit the technical details of the matching procedure.

\begin{figure}[t]
    \centering
    \includegraphics[width=\columnwidth]{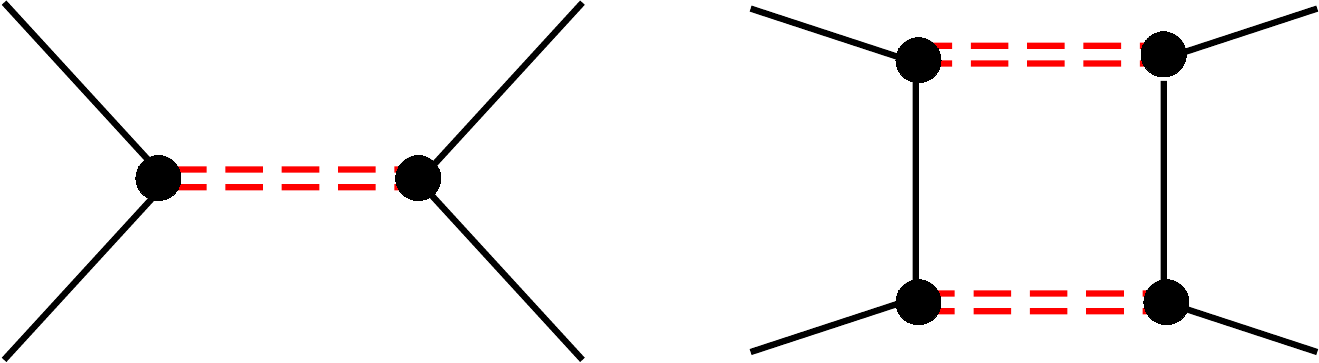}
    \caption{\it Tree-level and one-loop diagrams contributing to four-fermion interactions.}
    \label{fig:diagrams}
\end{figure}

The limited impact of finite one-loop matching corrections on lifting blind directions can be attributed to the fact that, in the limit of vanishing SM couplings, these corrections predominantly generate operators that lie within the same blind subspace defined in Eq.~\eqref{eq:blinddirections}. For instance, the scalar field $\Pi_1$ induces finite one-loop contributions to $\mathcal{O}_{ll}$ and $\mathcal{O}_{dd}$, in addition to the tree-level contribution to $\mathcal{O}_{ld}$, via diagrams such as those shown in Fig.~\ref{fig:diagrams}.

For higher-field extensions, we restrict our analysis to the scalar components, which can be consistently integrated out within existing one-loop matching frameworks. In what follows, we present explicit results for three illustrative scenarios.

\section{Fit for multi-field models}\label{sec:multi}
Finally, in this section, we fit three selected UV scenarios to the EWPOs, including the one-loop matching corrections from integrating out the heavy scalars involved.

\subsection*{Example 1}
We consider the following UV fields:
\begin{itemize}
    \item $\omega_1 \sim (3, 1)_{-1/3}$ (a scalar leptoquark)\,,
    \item $\mathcal{Q}_1 \sim (3, 1)_{1/6}$ (a vector leptoquark)\,.
\end{itemize}
The appearance of multiple leptoquarks is typical of GUTs and string-inspired models; see e.g. Ref.~\cite{Croon:2019kpe}. They are also common in models explaining flavour anomalies; see e.g. Refs.~\cite{Assad:2017iib,DAmico:2017mtc,Calibbi:2017qbu,Cornella:2019hct}.

\begin{figure}[t]
    \includegraphics[width=\columnwidth]{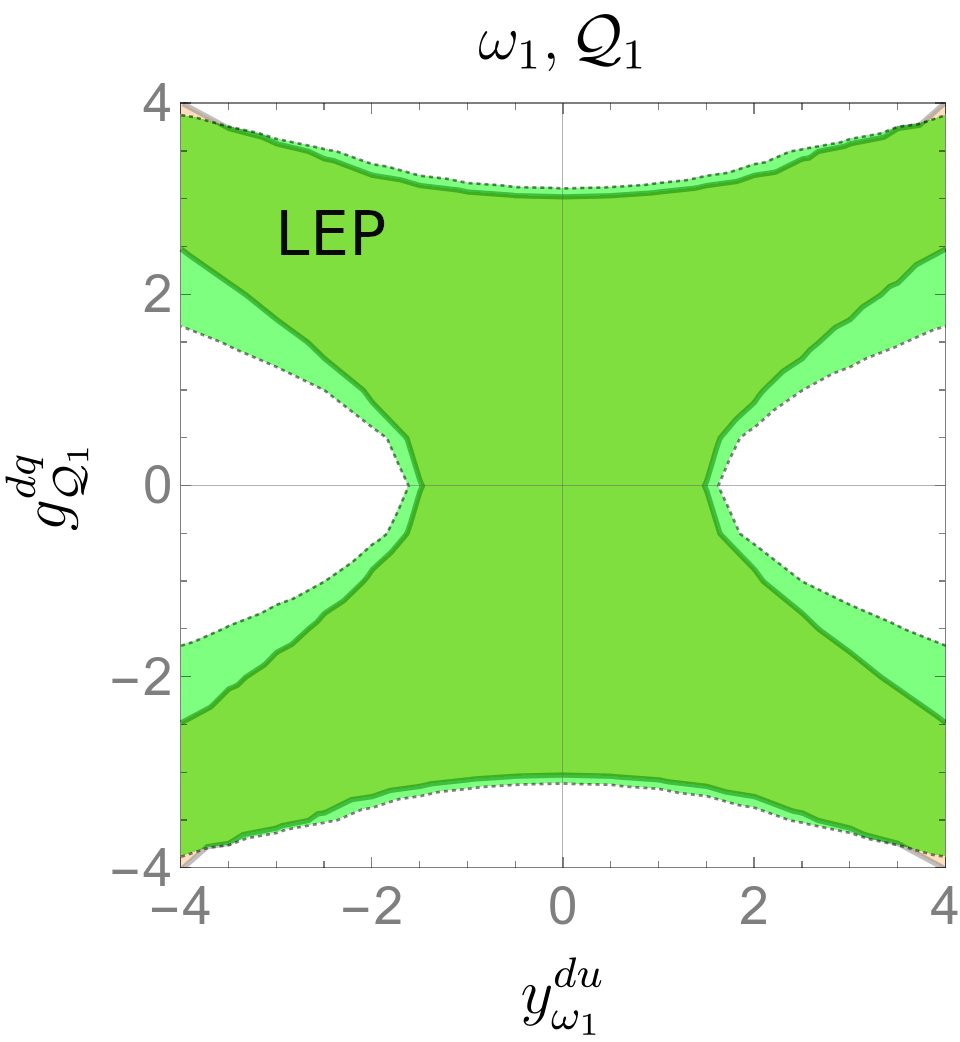}
    \caption{\it Region of the parameter space allowed by EWPOs at $1\,\sigma$ including tree-level and leading-log RGEs (dark area enclosed by solid line) and including one-loop matching scalar corrections and RGE resummation (light area enclosed by dashed line) for a UV model involving two leptoquarks $\omega_1$ and $\mathcal{Q}_1$. The masses of the scalar and the vector are assumed equal and of $1$ TeV.}
    \label{fig:uvmodels}
\end{figure}

The UV interaction Lagrangian reads:
\begin{align}
  \mathcal{L}_{UV} &=
  g^{ud}_{\omega_1} \epsilon_{ABC} \omega_1^{A\dagger} 
  \bar{d}^B_R u^{c\, C}_R
  \nonumber \\
  &\phantom{=}
  + y^{dq}_{\mathcal{Q}_1} \epsilon_{ABC} \mathcal{Q}_1^{A\mu\dagger}
  \bar{d}^B_R \gamma_\mu i \sigma_2 q^{c\, C}_L
  + \text{h.c.}\,.
\end{align}
Likewise, the effective Lagrangian is:
\begin{equation}
  \mathcal{L}_{\text{eff}}
  =
  \frac{|y^{du}_{\omega_1}|^2}{3 M_{\omega_1}^2}
  \left(\mathcal{O}^{(1)}_{ud} - 3 \mathcal{O}^{(8)}_{ud}\right)
  + \frac{2 |g^{dq}_{\mathcal{Q}_1}|^2}{3 M_{\mathcal{Q}_1}^{2}}
  \left(\mathcal{O}^{(1)}_{qd} - 3 \mathcal{O}^{(8)}_{qd}\right)\,.
\end{equation}
This gives rise to the following blind direction:
\begin{equation}
    \frac{|y^{du}_{\omega_1}|}{M_{\omega_1}}
    \simeq
    \sqrt{2} \frac{|g^{dq}_{\mathcal{Q}_1}|}{M_{\mathcal{Q}_1}}\,.
\end{equation}
The bounds imposed by EWPO on the parameter space of this model, obtained again with \texttt{smelli}, are shown in  Fig.~\ref{fig:uvmodels}. It is apparent that one-loop matching corrections only modify slightly the tree-level blind direction.

\begin{figure}[t]
    \includegraphics[width=\columnwidth]{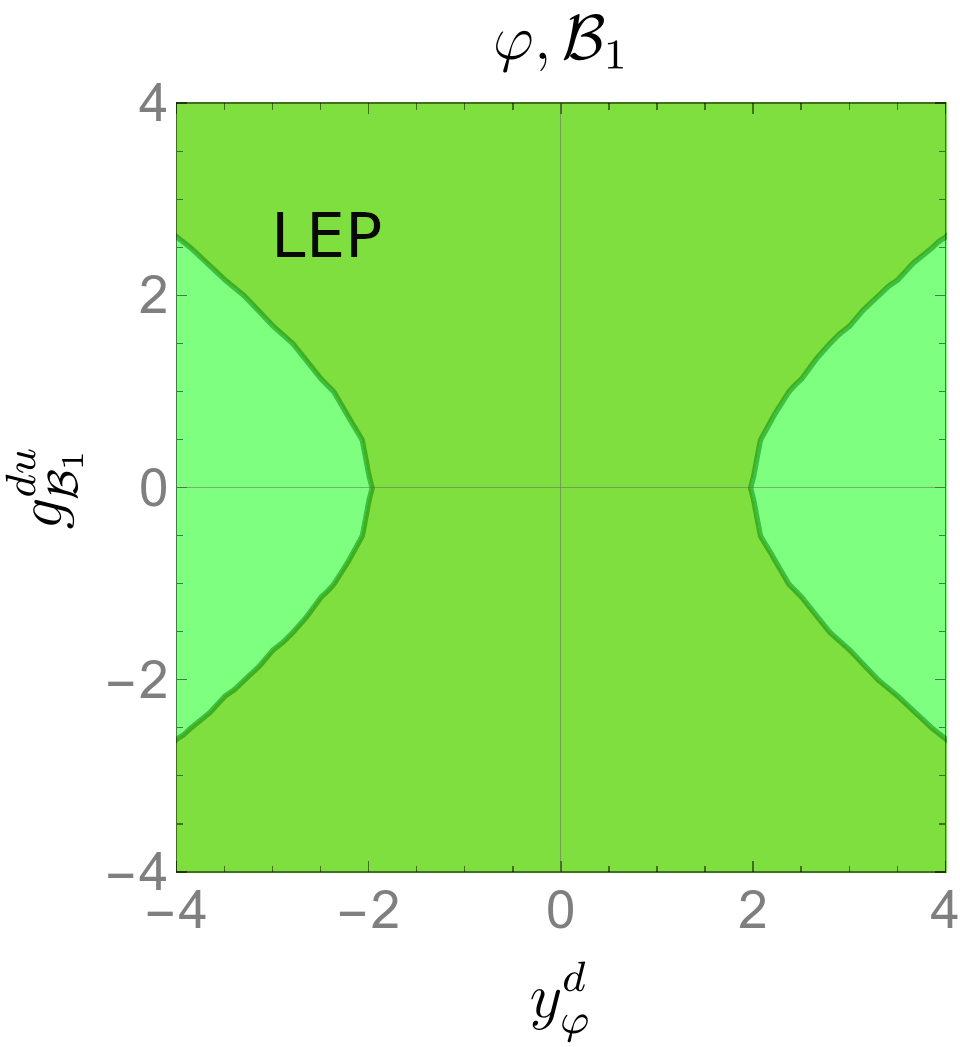}
    \caption{\it Same as Fig.~\ref{fig:uvmodels} but for an UV completion of the SMEFT involving $\varphi$ and $\mathcal{B}_1$.}
    \label{fig:uvmodels2}
\end{figure}

\subsection*{Example 2}
\label{sec:example-2}
Next, we consider the following UV fields:
\begin{itemize}
   \item $\varphi \sim (1, 2)_{1/2}$ (a second Higgs doublet)\,,
  \item $\mathcal{B}_1 \sim (1, 1)_1$ (a $W'$)\,.
\end{itemize}
Such a field content may be found in the left-right SUSY model, in which the SM gauge group is embedded into $SU(3)_C\times SU(2)_L\times SU(2)_R\times U(1)_{B-L}$. The $W'$ arises from the spontaneous breaking of $SU(2)_R$, while the second Higgs comes from the decomposition of Higgs bidoublets required by SUSY~\cite{delAguila:2010mx,Hirsch:2015fvq}.

The UV interaction Lagrangian reads:
\begin{equation}
  \mathcal{L}_{UV} =
    y^d_\varphi \varphi^\dagger \bar{d}_R q_L
    + g^{du}_{\mathcal{B}_1} \mathcal{B}^\dagger_{1\,\mu} \bar{d}_R \gamma_\mu u_R
  + \text{h.c.}\,.
\end{equation}
In turn, the effective Lagrangian is:
\begin{equation}
  \mathcal{L}_{\text{eff}}
  =
  - \frac{|y^d_{\varphi}|^2}{6 M_{\varphi}^{2}}
  \left(\mathcal{O}^{(1)}_{qu} + 6 \mathcal{O}^{(8)}_{qu}\right)
  - \frac{|g^{du}_{\mathcal{B}_1}|^2}{3 M_{\mathcal{B}_1}^{2}}
    \left(\mathcal{O}^{(1)}_{ud} + 6 \mathcal{O}^{(8)}_{ud}\right)\,.
\end{equation}
This gives rise to the following blind direction:
\begin{equation}
  \frac{|y^d_{\varphi}|}{M_{\varphi}} \simeq \sqrt{2} \frac{|g^{du}_{\mathcal{B}_1}|}{M_{\mathcal{B}_1}}\,.
\end{equation}
Using again \texttt{smelli}, we show the space constrained by EWPOs in Fig.~\ref{fig:uvmodels2}. Just like before, one-loop matching corrections are negligible.

\subsection*{Example 3}
\begin{figure}[t]
    \includegraphics[width=\columnwidth]{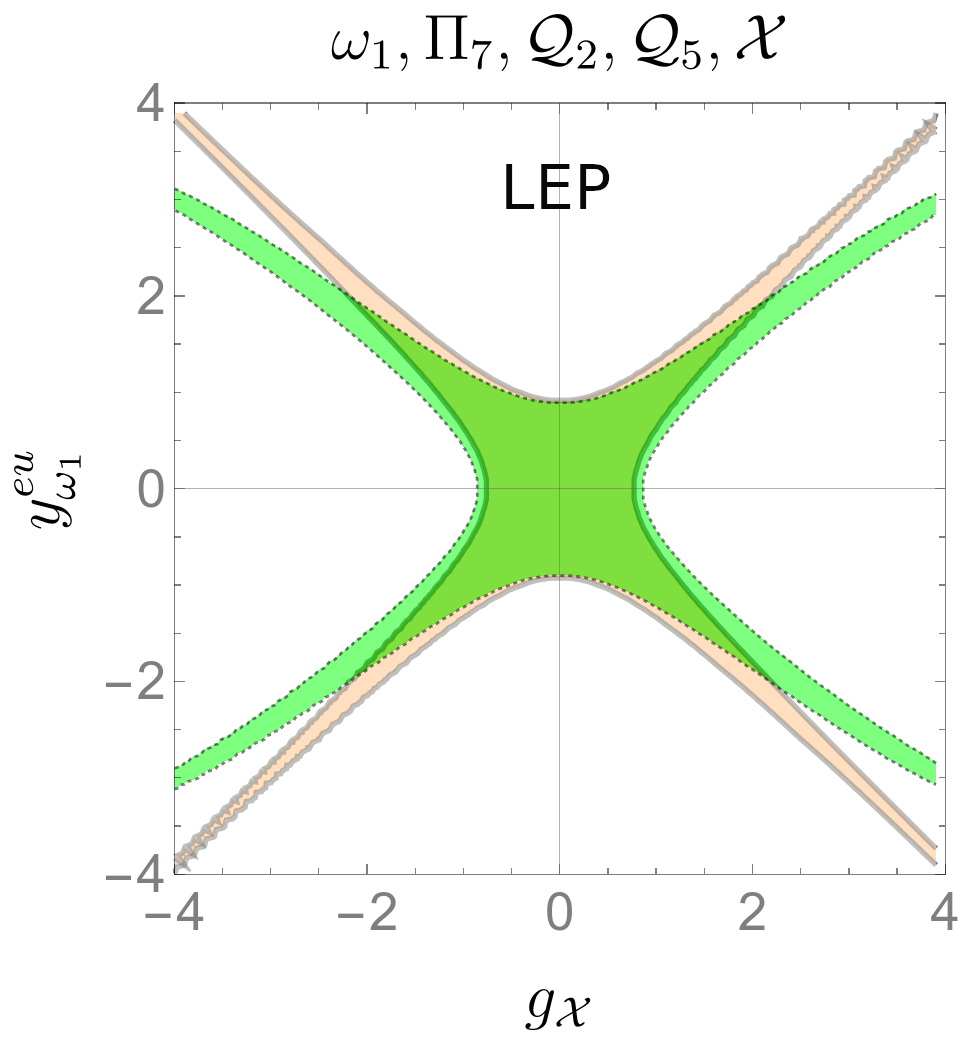}
    \caption{\it Same as Fig.~\ref{fig:uvmodels} but for an UV model with $\omega_1$, $\Pi_7$, $\mathcal{Q}_2$, $\mathcal{Q}_5$ and $\mathcal{X}$, assuming $\sqrt{2}g_{\mathcal{Q}_5}^{eq} = 2g_{\mathcal{U}_2}^{lq}=g_{\mathcal{X}}$ and $y_{\Pi_7}^{lu}=y_{\omega_1}^{eu}$.}\label{fig:uvmodels_3}
\end{figure}
As an example of a UV model that generates the blind direction $c_{eu} \sim c_{qe} \sim -c_{lu} \sim -c^{(1)}_{lq}$, we consider the following field content:
\begin{itemize}
    \item Scalars: $\omega_1 \sim (3, 1)_{-1/3}$ and $\Pi_7 \sim (3, 2)_{7/6}$\,,
    \item Vectors: $\mathcal{U}_2 \sim (3, 1)_{2/3}$, $\mathcal{Q}_5 \sim (3, 2)_{-5/6}$ and $\mathcal{X} \sim (3, 3)_{2/3}$\,,
\end{itemize}
with interaction Lagrangian
\begin{align}
    \mathcal{L}_{\text{UV}}
    &=
    y^{eu}_{\omega_1} \omega_1^\dagger \bar{e}^c_R u_R
    + g^{eq}_{\mathcal{Q}_5} \mathcal{Q}^{\mu\dagger}_5 \bar{e}^c_R \gamma_\mu q_L
    + y^{lu}_{\Pi_7} \Pi_7^\dagger i\sigma_2 \bar{l}^T_L u_R
    \nonumber \\
    &\phantom{=}
    + g^{lq}_{\mathcal{U}_2} \mathcal{U}^{\mu\dagger}_2 \bar{l}_L \gamma_\mu q_L
    + \frac{g_{\mathcal{X}}}{2} \mathcal{X}^{a\mu\dagger} \bar{l}_L \gamma_\mu \sigma^a q_L + \text{h.c.}\,.
\end{align}
All these heavy fields are leptoquarks, which arise in UV models pointed out in Example~2.
Integrating them out gives
\begin{align}
    \mathcal{L}_{\text{eff}}
    &=
    \frac{|y^{eu}_{\omega_1}|^2}{ 2 M_{\omega_1}^{2}} \mathcal{O}_{eu}
    + \frac{|g^{eq}_{\mathcal{Q}_5}|^2}{M_{\mathcal{Q}_5}^{2}}
       \mathcal{O}_{eq}
    - \frac{|y^{lu}_{\Pi_7}|^2}{ 2 M_{\Pi_7}^{2}} \mathcal{O}_{lu}
    \nonumber \\
    &\phantom{=}
    - \left(
        \frac{|g^{lq}_{\mathcal{U}_2}|^2}{2 M_{\mathcal{U}_2}^{2}}
        + \frac{3 |g_{\mathcal{X}}|^2}{8 M_{\mathcal{X}}^{2}}
      \right)
      \mathcal{O}^{(1)}_{lq}
    \nonumber \\
    &\phantom{=}
    - \left(
        \frac{|g^{lq}_{\mathcal{U}_2}|^2}{2 M_{\mathcal{U}_2}^{2}}
        - \frac{|g_{\mathcal{X}}|^2}{8 M_{\mathcal{X}}^{2}}
      \right)
      \mathcal{O}^{(3)}_{lq}\,.
\end{align}
The blind direction arises when
\begin{equation}
    \frac{|y^{eu}_{\omega_1}|}{M_{\omega_1}}
    \sim
    \sqrt{2} \frac{|g^{eq}_{\mathcal{Q}_5}|}{M_{\mathcal{Q}_5}}
    \sim
    \frac{|y^{lu}_{\Pi_7}|}{M_{\Pi_7}}
    \sim
    2 \frac{|g^{lq}_{\mathcal{U}_2}|}{M_{\mathcal{U}_2}}
    \sim
    \frac{|g_{\mathcal{X}}|}{M_{\mathcal{X}}}.
\end{equation}

The region of the parameter space constrained by EWPOs is shown in Fig.~\ref{fig:uvmodels_3}.

\section{\label{sec:terazcase} How things change at TeraZ}
TeraZ will measure EWPD with far greater accuracy than LEP, and will therefore be sensitive not only to effects driven by the large top Yukawa but also to RGE contributions induced by gauge couplings.

As an example, consider the parameter space spanned by ${c_{ud}^{(1)}, c_{qd}^{(1)}}$. EWPD constrain these operators through their mixing into $c_{\phi b}$ and $c_{\phi q}^{(1)}$, which directly affect the observables $A_{b}$ and $A_\text{FB}^b$:
\begin{align}
    \dot{c}_{\phi b} &\sim 6 y_t^2 (c_{qd}^{(1)}-c_{ud}^{(1)})+\frac{2}{3}g_1^2 (c_{qd}^{(1)}+2 c_{ud}^{(1)})\,,\\
    \dot{c}_{\phi q}^{(1)} &\sim -\frac{2}{3} g_1^2 c_{qd}^{(1)}\,.
\end{align}

\begin{figure}[t]
\includegraphics[width=0.49\columnwidth]{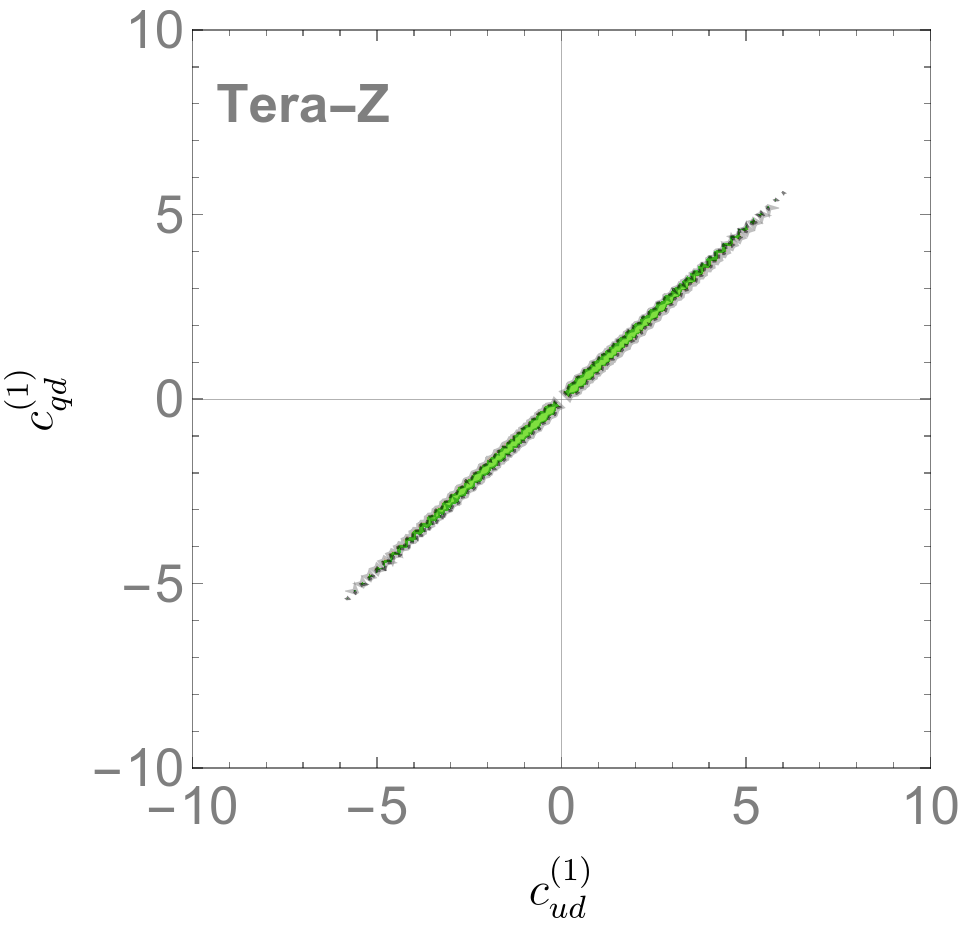}
\includegraphics[width=0.49\columnwidth]{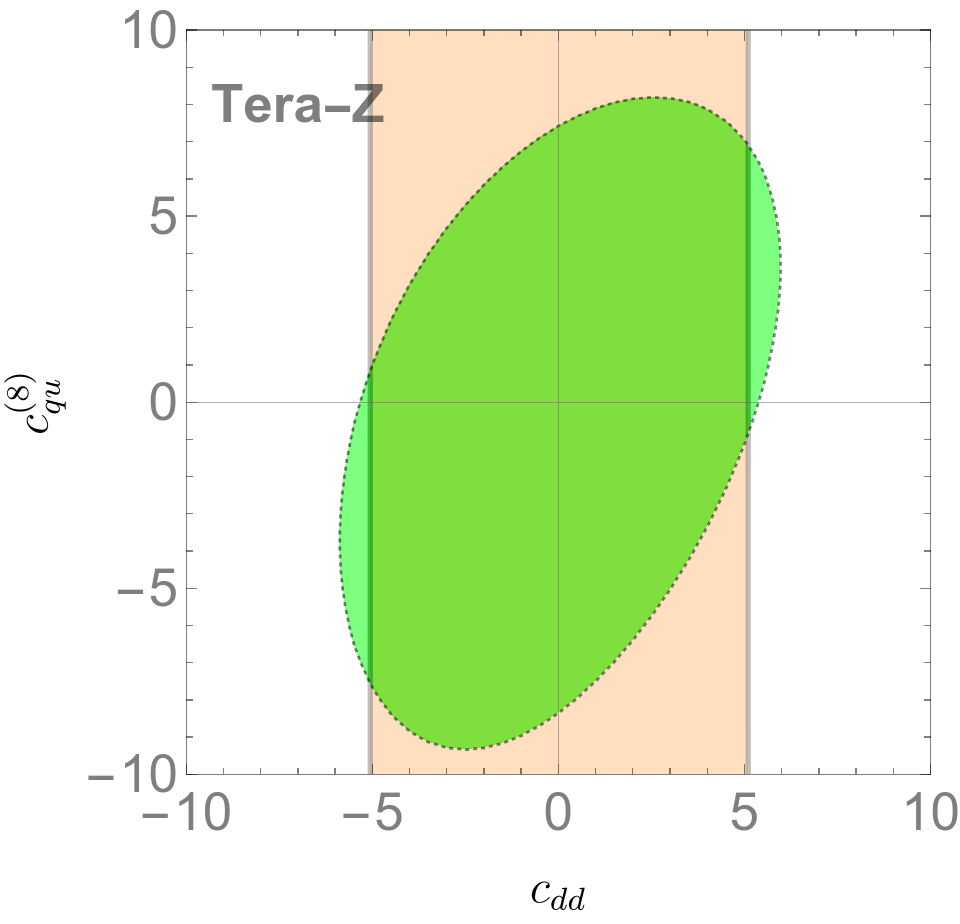}
\includegraphics[width=0.49\columnwidth]{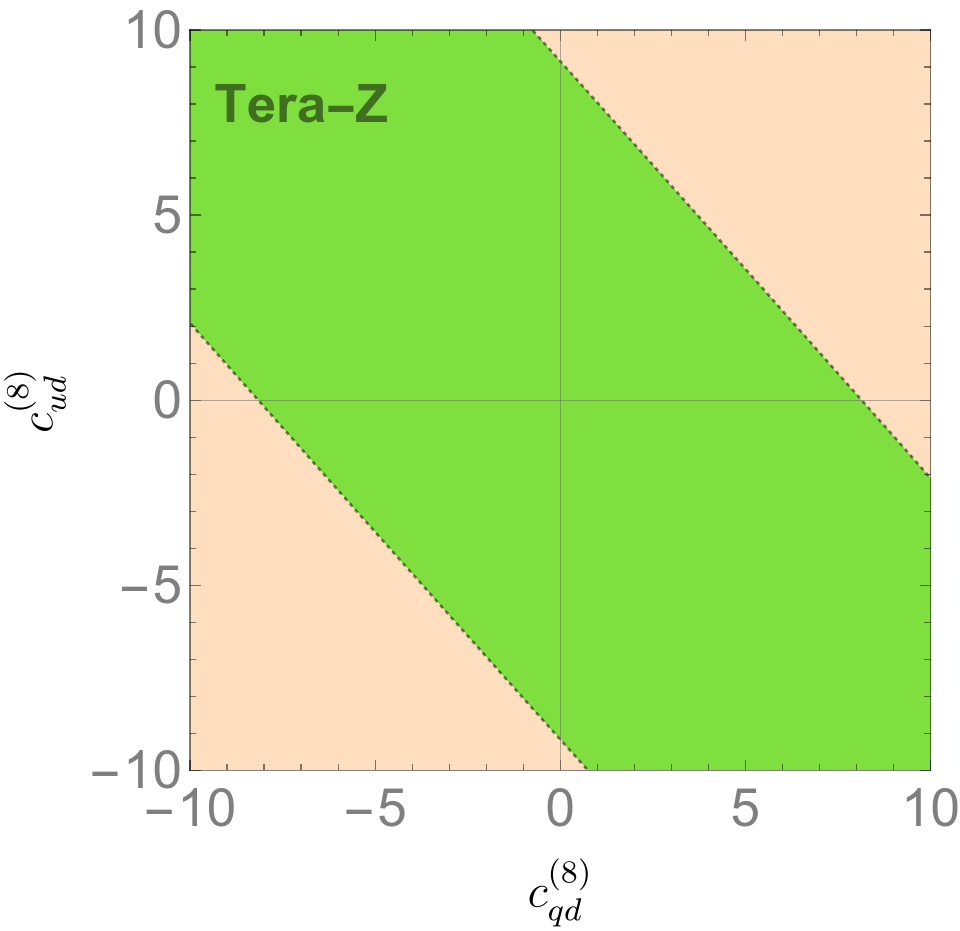}
\includegraphics[width=0.49\columnwidth]{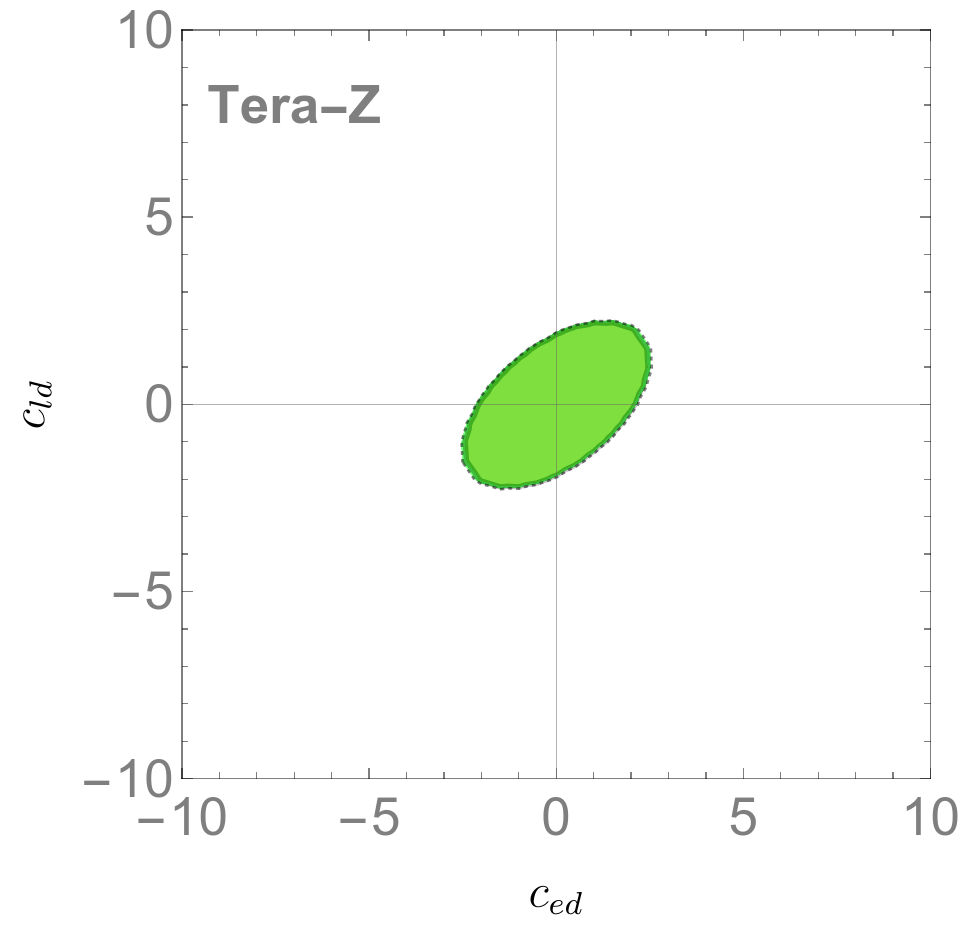}
\caption{\it Fate of some of the LEP blind directions at TeraZ. Colour-code is as in Fig.~\ref{fig:blinddirections}. We have used the modified versions of \texttt{smelli} and \texttt{flavio} presented in Ref.~\cite{Allanach:2025wfi}.}\label{fig:blinddirectionteraZ}
\end{figure}

At LEP precision, $g_1^2$ contributions can be neglected, which leads to the blind direction $c_{qd}^{(1)} \sim c_{ud}^{(1)}$. At TeraZ, however, $g_1^2$ effects are no longer negligible. Nevertheless, a blind direction persists, understood here as a region of parameter space where TeraZ sensitivity drops very significantly, and the same holds for most of the \textit{basic} blind directions identified above;\footnote{Note that, beyond these basic ones, exact blind directions will always exist, given that the parameter space has larger dimensionality than the number of EWPO. From an agnostic EFT standpoint, without additional knowledge about the UV, they cannot be simply dismissed as unlikely.} see Fig.~\ref{fig:blinddirectionteraZ}.

\begin{figure}[t]
 \includegraphics[width=\columnwidth]{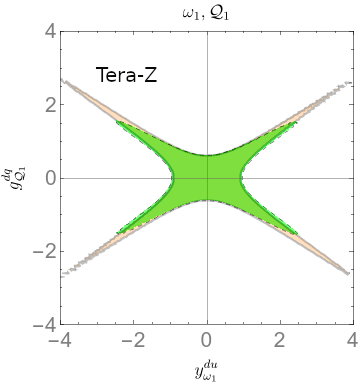}
 \caption{\it TeraZ constraints at $2$ $\sigma$ on the model of Fig.~\ref{fig:uvmodels_3} using the same colour-coding.}\label{fig:terazmodel}
\end{figure}

The only ``exceptions" are the blind directions involving $c_{qe}, c_{lq}^{(1)}, c_{eu}, c_{lu}$, not shown in the plots, for which couplings above $\sim 0.5$ can already be excluded at TeraZ. For all other cases, the blind directions merely narrow, and even couplings as large as $g \sim 0.5$ would remain beyond TeraZ sensitivity. This is also reflected in the fact that the simple SM extensions constructed earlier remain largely unconstrained by TeraZ data; see, for instance, Fig.~\ref{fig:terazmodel}.

One final comment is in order. Although our analysis assumes new physics at the TeV scale, running down to the electroweak scale via RGE, it is reasonable to expect that if such large couplings remain unconstrained, then new physics well below the TeV scale with $  g\lesssim 1$ could also evade detection at TeraZ.

This limitation arises from working within EFT at a fixed scale, which is sensitive only to the ratio $g/\Lambda$ but not to $g$ and $\Lambda$ separately. This stands in contrast to hadron colliders. For instance, the particles in Fig.~\ref{fig:terazmodel} could be searched for pair-produced dijet resonances~\cite{CMS:2024ldy} via gluon fusion, such as $gg \to \omega_1 \omega_1$ with $\omega_1 \to jj$. The small coupling to quarks, undetectable at TeraZ, would merely increase the $\omega_1$ lifetime, without otherwise diminishing the sensitivity of hadron collider searches. This further motivates later high-energy stages of the FCC programme, including $e^+e^-$ runs at different energies.

\section{\label{sec:summary} Summary and Conclusions}
In this work, we have investigated blind directions in the SMEFT, focusing on the four-fermion sector with third-family WCs. By restricting our analysis to this sector, we have identified 
high-dimensional subspace that remains blind to EWPOs, even after accounting for improved perturbation theory. We have shown that simple multi-field UV completions of the SM, where the heavy fields are integrated out at tree level, can reside within this blind subspace. 
One-loop matching corrections do not alter the persistence of these blind directions.

Our results make it clear that blind directions are not isolated curiosities but rather a systematic and generic feature of realistic SMEFT completions. Even under simplifying assumptions, such as limiting to third-generation couplings, working with tree-level matching, and considering only a modest number of operators, we identify multiple blind directions that evade the entire suite of EWPOs at TeraZ. The situation becomes even more pronounced when considering that 25 independent four-fermion operators (which can all be generated at tree level) are constrained by only 26 EWPOs, many of which are sensitive to overlapping subsets of operators. 
Our findings reveal that the problem is structural rather than incidental,
highlighting limitations of relying solely on precision measurements, even at the unprecedented luminosities of TeraZ, to comprehensively probe new physics.

These results provide further motivation for complementary high-energy collider experiments. Future FCC-ee runs at higher center-of-mass energies and the FCC-hh will open access to new kinematic regimes and help resolve the degeneracies that remain in precision electroweak measurements. Their role becomes especially valuable as the typical scale of new physics might extend beyond the 1 TeV benchmark considered in our analysis.


\begin{acknowledgments}
This work has received funding from MICIU/AEI/10.13039/501100011033 and ERDF/EU
(grants PID2022-139466NB-C22 and PID2021-128396NB-I00), from the Junta de Andaluc\'ia grants FQM 101 and P21-00199 as well as from The Royal Society programmes Newton International Fellowship Alumni follow-on and International Exchanges under grant numbers AL-231018 and IES-R1-221239, respectively. MC and JCC are further supported by the Ram\'on y Cajal program under grants RYC2019-027155-I and RYC2021-030842-I; respectively.
\end{acknowledgments}


%

\bibliography{references}

\end{document}